# Music Enhances Activity in the Hypothalamus, Brainstem, and Anterior Cerebellum during Script-Driven Imagery of Affective Scenes

**Chia-Wei Li[1], Tzu-Han Cheng[2], and Chen-Gia Tsai [3,4]***

[1] Department of Radiology, Wan Fang Hospital, Taipei Medical University, Taipei, Taiwan

[2] Neurobiology and Cognitive Science Center, National Taiwan University, Taipei, Taiwan

[3] Department of Cognitive Science, UC San Diego, La Jolla, CA, USA

[4] Graduate Institute of Musicology, National Taiwan University, Taipei, Taiwan

**\* Please send correspondence to:**

Chen-Gia Tsai, Ph.D.

Graduate Institute of Musicology, National Taiwan University, No.1, Sec. 4, Roosevelt Road, Taipei, 106, Taiwan. (R.O.C.)
Email: tsaichengia@ntu.edu.tw

Abstract

Music is frequently used to establish atmosphere and to enhance/alter emotion in dramas and films. During music listening, visual imagery is a common mechanism underlying emotion induction. The present functional magnetic resonance imaging (fMRI) study examined the neural substrates of the emotional processing of music and imagined scene. A factorial design was used with factors emotion valence (positive; negative) and music (withoutMUSIC: script-driven imagery of emotional scenes; withMUSIC: script-driven imagery of emotional scenes and simultaneously listening to affectively congruent music). The baseline condition was imagery of neutral scenes in the absence of music. Eleven females and five males participated in this fMRI study. Behavioral data revealed that during scene imagery, participants' subjective emotions were significantly intensified by music. The contrasts of positive and negative withoutMUSIC conditions minus the baseline (imagery of neutral scenes) showed no significant activation. When comparing the withMUSIC to withoutMUSIC conditions, activity in a number of emotion-related regions was observed, including the temporal pole (TP), amygdala, hippocampus, hypothalamus, anterior ventral tegmental area (VTA), locus coeruleus, and anterior cerebellum. We hypothesized that the TP may integrate music and the imagined scene to extract socioemotional significance, initiating the subcortical structures to generate subjective feelings and bodily responses. For the withMUSIC conditions, negative emotions were associated with enhanced activation in the posterior VTA compared to positive emotions. Our findings replicated and extended previous research which suggests that different subregions of the VTA are sensitive to rewarding and aversive stimuli. Taken together, this study suggests that emotional music embedded in an imagined scenario is a salient social signal that prompts preparation of approach/avoidance behaviours and emotional responses in listeners.

# 1. Introduction

There are multifaceted relationships between music, emotion, and scenes. Compositions of classical *program music* such as Beethoven's *Pastorale* symphony are based on specific scenes that evoked composer's feelings and emotions. Affective-emotional appreciation of music is sometimes accompanied by voluntarily or involuntarily imagined scenes. The imagined scenes may help bringing real-world relevance to music, thereby augmenting the emotional processing of music. This view is supported by the finding that emotional music on its own did not yield significant activity increase in the amygdala and hippocampus, but their activity significantly increased when the same music was combined with neutral films (Eldar et al., 2007).

In addition to the scene-to-music influences, there are also music-to-scene influences. Music is frequently used to establish atmosphere and to enhance/alter emotion in dramas and films. Evidence has accumulated to show that the interaction between the superior temporal gyrus (STG), temporal pole (TP), and subcortical regions is essential for the music-to-scene influences. Baumgartner et al. (2006) observed that music enhanced subjects' feeling of emotional pictures and increased activity in the amygdala, hippocampus, striatum, cerebellum, fusiform, and STG. Pehrs et al. (2014) investigated how music impacted the processing of dynamic kissing scenes, finding that the fusiform to amygdala coupling strength was modulated by the anterior STG. Using empathy-evoking film clips as stimuli, Pehrs et al. (2017) recently showed that emotional music enhanced the TP to fusiform connectivity during the integration of contextual information, suggesting that the TP integrates multimodal emotional information and top-down modulates the ventral visual stream.

Although previous studies have explored how music enhances the emotional responses to pictures or films, no functional magnetic resonance imaging (fMRI) experiment has examined the effects of music on scene imagery. In the present study, we employed fMRI and a task of script-driven scene imagery to examine the neural correlates underlying the emotional processing of music and imagined scenes. The issue of interaction between emotional music and scene imagery is important because visual imagery is one of the six mechanisms underlying emotion induction via music (Juslin and Vastfjall, 2008). A study on the affective effects of extramusical information about music indicated that narrative descriptions enhanced emotion induction via the visual imagery mechanism (Vuoskoski & Eerola, 2015). Moreover, combinations of visual imagery and music listening are found to have therapeutic benefit to various affective disturbances (Karagozoglu *et al.*, 2013; McKinney & Honig, 2017).

The nature of subjective experience seems to differ between scene imagery and picture/film perception. Unlike picture/film perception that is based on input visual information, scene imagery involves the generation, maintenance, and visualization of an environmental setting into which multimodal details of an experience are bound (Burgess *et al.*, 2001; Hassabis *et al.*, 2007b). When a musical excerpt is delivered along with a picture or a video clip, a perceiver may try to obtain a better understanding of this scene via multimodal integration. On the other hand, when one is imagining a scene during exposure to a musical excerpt, active construction of an environmental setting is likely affected by the music, which could help the imaginer vividly experiencing an event. We hypothesized that during scene imagery, emotional music may cause a shift in perspective taking. When the participants were instructed to imagine a fictitious emotional scene in the absence of music, they tended to take a cognitive perspective and keep a psychological distance apart from the imagined persons and scenes. When the participants imagined this scene with a simultaneous presentation of affectively congruent music, they may take an affective perspective, re-enacting the protagonist's emotional responses and feelings, because this music represented the protagonist's emotions, feelings, and emotion-related movements/vocalizations (Molnar-Szakacs & Overy, 2006; Panksepp, 2009). Previous studies suggest that music with a specific emotional expression is likely to elicit the same emotion in the listener (Lundqvist *et al.*, 2009; Hunter *et al.*, 2010), and emotion contagion is a prominent mechanism underlying emotion induction via music (Juslin and Vastfjall, 2008). The major aim of this study was to examine how music altered activity in emotion-related regions during scene imagery. Based on the aforementioned study by Baumgartner et al. (2006), we predicted that a simultaneous presentation of affectively congruent music during scene imagery may significantly enhance activity in the amygdala, hippocampus, and cerebellum.

  The ability to recognize and to re-enact the emotions expressed by music varies across individuals. Trait empathy, defined as the ability to understand and share affective responses/states experienced by other individuals (Mehrabian & Epstein, 1972), was found to have modulatory effects on the subjective emotions induced by music (Egermann & McAdams, 2013). In the present study, volunteers first took an online pre-scan test, in which they were asked to establish one-to-one association between ten emotional music excerpts and ten linguistic descriptions of emotional scenes. Those who reported consistent recognized affect contents conveyed by the musical stimuli were allowed to proceed to the fMRI experiment. Given that cognitive empathy has profound impacts on affective physiological responses to music (Miu & Baltes, 2012), emotional music embedded in an imagined scenario may induce subjective emotional feelings and bodily reactions in these "empathic" participants. We expected that imagery of affective scenes and simultaneously listening to congruent music may be associated with significantly increased activity in the hypothalamus and brainstem, relative to pure imagery of affective scenes. This study promises to yield novel data regarding how music initiates the ancient subcortical structures to generate emotional responses.

## 2. Methods

### 2.1 Participants

General advertising online and in local media was used to recruit participants. Sixty-six participants took an online pre-scan test, in which the participants were asked to establish one-to-one association between ten emotional music excerpts and ten linguistic descriptions of emotional scenes (see Stimuli and Procedure). Twenty-three of them had correct rates of 80% or above in the test, and therefore were invited to participate in the fMRI experiment. Imaging data of sixteen participants (mean age 22.44±1.58 years; 11 females) were collected and analyzed. All of them were free of any neurological, psychiatric, or hearing disorders. Each participant gave written informed consent to participate in the study, and received monetary compensation for participation. The study procedures were conducted in accordance with the Declaration of Helsinki and approved by the Research Ethics Committee of National Taiwan University.

### 2.2 Stimuli

The musical stimuli of this study were ten music excerpts. Each of them was 15 seconds in length. Five of them expressed positive emotions such as happiness and fun; five of them expressed negative emotions such as sadness, fear, and despair. Our main criterion for stimulus selection was maximization of the intensity of music-induced emotions. Eight music excerpts were selected from a corpora of music-related emotions established in Taiwan (Chen, 2013); three excerpts conveyed strong positive emotions and five excerpts conveyed strong negative emotions from this corpora. These eight excerpts have been rated as conveying the highest or the second highest emotion intensity in positive or negative emotion categories in a previous study (Chen, 2013). In addition, we selected two funny music excerpts as positive stimuli from two websites (http://audionautix.com/ and http://www.purple-planet.com/). These two funny music excerpts were expected to induce strong positive emotion. Musical instruments and styles in all ten excerpts ranged from Western classical music, Chinese classical music, to popular music. A female humming voice appeared in one excerpt (positive emotion), and a male shouting voice appeared in another excerpt (negative emotion). No musical stimuli contained comprehensible lyrics.

Linguistic stimuli in Mandarin were fifteen scenarios created by the first author. They served as the scripts guiding the participants to imagine specific scenes. Five linguistic stimuli depicted the scenes corresponding to the five positive music excerpts and five linguistic stimuli depicted the scenes corresponding to the five negative music excerpts (https://youtu.be/EQ2IlOUhLRc). The other five linguistic stimuli depicted scenes with neutral emotion. The length of these linguistic stimuli ranged from 19 to 25 characters. The complete list of linguistic stimuli (translated into English) is given in the Table 1.

2.3 Procedure

Participant screening included an online pre-scan test, in which the participant was asked to establish one-to-one association between ten emotional music excerpts and ten linguistic stimuli depicting emotional scenes. The inclusion criterion for fMRI scans was 80% correct answers. The correct answers were defined as the mostly chosen associations. If more than 33 % participants of this online test chose a linguistic stimulus for a certain musical stimuli, this association was regarded as correct. The purpose of this screening was to enroll a group reporting relatively consistent recognized affect contents conveyed by the musical stimuli.

A schematic diagram of the procedure is shown in Figure 1. There were five runs in the fMRI scan. Each run contained 15 trials. Each trial had a duration of 24 s. It required 360 seconds to finish each run and 30 minutes to finish the whole fMRI scan. In each trial, the participants were instructed to visually imagine the scene described by the script (linguistic stimulus) shown on the screen. Meanwhile, a musical stimulus or silence was presented for 15 seconds. At the end of each trial, an instruction was presented on the screen, which asked the participant to report their present subjective emotion valence by pressing one of the two response buttons (left button for "negative" and right button for "positive"). If the participant felt neutral emotion, no button-pressing was required.

We used a 2 (positive or negative valence) by 2 (withMUSIC or withoutMUSIC) factorial design to test our hypothesis that emotional music may cause a shift in perspective taking during scene imagery. There were four experimental conditions (positive withMUSIC, negative withMUSIC, positive withoutMUSIC, and negative withoutMUSIC) and a baseline condition. The baseline condition was imagery of neutral scenes without music. For each condition, there were five scenes described by linguistic stimuli. Within each condition, each linguistic stimulus repeated for three times throughout the fMRI scan. In total, there were 75 trials presented in a pseudo-random order.

2.4 MRI data acquisition

For imaging data collection, participants were scanned using a 3T MR system (MAGNETOM Prisma, Siemens, Erlangen, Germany) and a 20-channels array head coil at the Imaging Center for Integrated Body, Mind, and Culture Research, National Taiwan University. In the functional scanning, about 2.5 mm slices of axial images were acquired using a gradient echo planar imaging with the following parameters: time to repetition = 2500 ms, echo time = 30 ms, flip angle = 87°, in-plane field of view = 192 × 192 mm, and acquisition matrix = 78 × 78 × 45 to cover whole cerebral areas. For spatial individual-to-template normalization in preprocessing, a Magnetization Prepared Rapid Gradient Echo T1-weighted imaging with spatial resolution of 0.9 mm isotropic was acquired among each participant.

2.5 Data processing and statistical analysis

Preprocessing of fMRI data was performed with SPM12 (Wellcome Trust Centre for Neuroimaging; http://www.fil.ion.ucl.ac.uk/spm) and the Artifact Detection Tools (ART) toolkit. The data preprocess included 6 steps. First, the first four volumes of each run were discarded to allow for magnetic saturation effects. Second, the remaining functional images were corrected for timing differences in slice acquisitions. Image slices were acquired in interleaved order, and the slice-timing correcting was conducted with the use of the first slice of each scan as the reference. Third, we removed the slice-timing corrected dataset from head movement artifact using rigid-body realignment in SPM12. Fourth, the preprocessed functional images were then coregistered to the individual anatomical image, and it was subsequently used to normalize the coregistered functional images into the standard MNI brain template with 2-mm isotropic voxel size. Fifth, the normalized images were spatially smoothed with a Gaussian kernel of 5-mm full width at half maximum (FWHM) to accommodate any anatomical variability across participants. In the final step of data preprocess, the ART toolkit was used to check the spikes and motion of each dataset. All functional dataset showed head motion below 2.5 mm of maximal translation (in any direction) and 1° of maximal rotation throughout the course of scanning, and none outliers happened in all functional dataset.

Analysis of imaging data was performed using SPM12. A 2 (positive valence; negative valence) by 2 (withoutMUSIC; withMUSIC) repeated-measures ANOVA and subsequent Bonferroni post-hoc tests were performed on the self-rating scores of subjective emotion valence. Statistical inference was based on a random effect approach at two levels. At the single-subject level, the data of each participant were analyzed using the general linear model by fitting the time series data with the canonical hemodynamic response function (HRF) modeled at events (Josephs & Henson, 1999). The events were modeled for the presentation of music or silence in each of the five conditions (Figure 1). Linear contrasts were computed to characterize responses of interest, averaging across fMRI runs. The group-level analysis consisted of paired *t*-tests for the contrasts of four experimental conditions minus the baseline condition, a 2 × 2 ANOVA with factors emotion valence and music, and subsequent post-hoc paired-sample *t*-tests. Statistical significance was thresholded at false discovery rate (FDR) of $p < 0.05$ at the cluster level, with a minimum cluster size of 10 voxels.

3. Results

Figure 2 presents the behavioural data of the self-rating scores of subjective emotion valence (-1: negative; 0: neutral, 1: positive). The two-way ANOVA revealed a significant main effect of factor emotion valence ($F(1, 15) = 182.921$, $p < 0.001$) and factor music ($F(1, 15) = 5.404$, $p = 0.035$). The rating scores of emotion valence were higher for positive-emotion conditions than for negative-emotion conditions. The rating scores of emotion valence were higher for withMUSIC conditions than for withoutMUSIC conditions. There was a significant interaction effect ($F(1, 15) = 78.334$, $p < 0.001$). Post-hoc Bonferroni-corrected pairwise comparisons revealed that the scores of the positive withMUSIC condition were significantly higher than those of the positive withoutMUSIC condition; the scores of the negative withMUSIC condition were significantly lower than those of the negative withoutMUSIC condition. These results indicated that emotional music significantly increased the subjective emotion intensity of imagined emotional scenes.

The group-level fMRI results of the contrasts of four experimental conditions minus the baseline are shown in Table 2 and Figure 3. The transverse temporal gyrus, insula, STG, TP, amygdala, hippocampus, locus coeruleus (LC), cerebellum, and hypothalamus were commonly activated for the positive and negative withMUSIC conditions relative to the baseline condition.

The results of the two-way ANOVA of fMRI data are summarized in Table 3. There was a main effect of music. Compared to the withoutMUSIC conditions, the withMUSIC conditions were associated with greater activity in the premotor cortex, STG, insula, temporal pole, amygdala, hippocampus, thalamus, hypothalamus, LC, midbrain, anterior ventral tegmental area (VTA), and anterior cerebellum (Figure 4). There was also a main effect of emotion valence. Compared to the negative emotion conditions, the positive emotion conditions were associated with greater activity in the STG, insula, cingulate cortex, precuneus/cuneus, occipital gyrus, inferior parietal gyrus, supramarginal gyrus, fusiform gyrus, inferior temporal gyrus, and medial orbitofrontal cortex. Moreover, significance was found in the STG, posterior insula, and posterior VTA for the interaction effect. For the withMUSIC conditions, positive emotions were associated with enhanced activation in the posterior insula and reduced activation in the posterior VTA compared to negative emotions (Figure 5).

## 4. Discussion

The present study aimed at examining the neural correlates of the emotional processing of scene imagery and affectively congruent music. We selected participants reporting consistent recognized affect contents conveyed by the musical stimuli for the fMRI experiment. While undergoing fMRI, the participants were asked to imagine scenes depicted by the script in the presence or absence of congruent music. As expected, subjective ratings of emotional valence for the positive withMUSIC condition were significantly higher than those for the positive withoutMUSIC condition, while subjective ratings of emotional valence for the negative withMUSIC condition were significantly lower than those for the negative withoutMUSIC condition. We found that neural activation associated with the positive and negative withoutMUSIC conditions did not significantly differ from that associated with the baseline (imagery of neutral scenes in silence). When comparing the withMUSIC conditions to withoutMUSIC conditions, activity in the bilateral TP and several subcortical emotion-generative systems was observed. In particular, we demonstrated that activity in the amygdala, hippocampus, hypothalamus, anterior VTA, LC, and anterior cerebellum was significantly enhanced by music. Moreover, the two-way ANOVA revealed a significant interaction effect between emotion valence and music. For the withMUSIC conditions, the contrast of negative minus positive emotions yielded activation in the posterior VTA. These findings indicated that emotional music embedded in an imagined scene initiated the ancient subcortical structures to generate emotional responses.

The contrast of the withMUSIC minus withoutMUSIC conditions showed activity in the TP, amygdala, and hippocampus. It is well established that the anterior temporal lobe plays a central role in socioemotional processing and empathy (Royet *et al.*, 2000; Leigh *et al.*, 2013; Olson *et al.*, 2013). With regard to musical emotions, past studies of patients with dementia have indicated that the inability to recognize emotions conveyed by music was associated with impairment of the anterior temporal lobe (Omar *et al.*, 2011; Hsieh *et al.*, 2012; Downey *et al.*, 2013). Differences in function between TP subregions have been investigated by examining their anatomical and functional connection patterns. The dorsal TP is mainly connected with the superior temporal areas, inferior frontal gyrus, and insular cortex, whereas ventral TP is connected with the association visual cortex in the inferior temporal cortex, amaygdala, hippocampus, and orbitofrontal cortex (Olson *et al.*, 2013; Fan *et al.*, 2014; Murphy *et al.*, 2017). In the present study, the activated TP clusters for the contrast of withMUSIC > withoutMUSIC comprised the dorsal and ventral subregions of the bilateral TP. Based on the aforementioned studies, we speculate that during script-driven scene imagery and simultaneously listening to music, the dorsal TP was mainly responsible for the auditory and semantic processing, whereas the ventral TP may integrate the inputs from the dorsal TP with the imagined visual scene to extract socioemotional significance, and then regulated the subcortical emotion-generative systems. This view is consistent with previous observations that the TP and anterior STG contributed to integrative socioemotional processes of scene-music combinations via gating the coupling between the fusiform and amygdala (Pehrs *et al.*, 2014; Pehrs *et al.*, 2017).

Having connections to cortical and subcortical regions, the amygdala and hippocampus are widely recognized as hubs linking cognitive and emotional processes (Robinson *et al.*, 2012; Bickart *et al.*, 2014; Backus *et al.*, 2016; Mears & Pollard, 2016). In the present study, we speculate that the ventral TP exerted top-down modulations to the amygdala during exposure to emotional music. Moreover, we observed coactivation of the amygdala and thalamus for the contrast of withMUSIC > withoutMUSIC. This finding is partially in accordance with research by Koelsch and Skouras (2014), who reported increased functional connectivity between the amygdala and the thalamus during exposure to joyful music.

The amygdala and hippocampus recruit appropriate autonomic and endocrine responses through their projections to hypothalamic-brainstem structures (Ulrich-Lai & Herman, 2009). Prior evidence has suggested that both positively- and negatively-valenced stimuli trigger these responses. Koelsch and Skouras (2014) reported increased functional connectivity between the hippocampus and hypothalamus during exposure to joyful music. Lerner et al. (2009) demonstrated that increased amygdala activation in response to negatively-valenced music with eyes closed was associated with increased activations in the LC, a key brainstem region for noradrenergic modulation of arousal and attention (Waterhouse & Navarra, 2019). The hypothalamic neurons that produce hypocretins send projections to several midbrain regions, including the LC and VTA (Li *et al.*, 2016). This may explain the coactivation of the hypothalamus, LC, and anterior VTA for the contrast of withMUSIC > withoutMUSIC. The VTA is a core component of the dopamine reward system. A few studies have reported its sensitivity to pleasurable music (Menon & Levitin, 2005; Trost *et al.*, 2012; Li *et al.*, 2015).

The posterior VTA exhibited greater activity for the negative than positive withMUSIC conditions. Engen et al. (2017) recently found that the posterior part of the midbrain tracked the intensity of negative affect, although that region was more posterior than the posterior VTA found in the current study. Emerging evidence suggests that both rewarding and aversive stimuli activate VTA, and this is partially due to the functional and anatomical heterogeneity of the VTA (Holly & Miczek, 2016). Future research is needed to elucidate the functional connectivity and neurotransmitter release profiles within the VTA during processing the positive and negative emotions of scene-music combinations.

Compared to the withoutMUSIC conditions, the withMUSIC conditions were associated with significantly increased activation in the anterior lobule of the cerebellum. To the best of our knowledge, there has been only one fMRI study by Baumgartner et al. (2006) showing anterior cerebellum activation in response to emotional music, and their experimental design resembled to that of the current study. Baumgartner et al. (2006) presented fearful and sad pictures either alone or combined with congruent emotional music excerpts, finding that music elicited activation in the anterior cerebellum. Here we replicated their finding using the task of script-driven scene imagery. The exact role of the anterior cerebellum in the emotional processing of scene-music combinations is unclear. Baumgartner et al. (2006) discussed their finding in relation to a hypothesis made by Damasio et al. (2000) that the cerebellum may modulate/coordinate varied emotional action programs. We speculate that the anterior cerebellum might contribute to preparation of approach and avoidance behaviours related to positive and negative scene-music combinations, respectively. This view echoes research by Eldar et al. (2007), who suggested that subcortical responses to the combinations of emotional music and neutral film clips may be driven by the emotion-related content of a stimulus that determines the goal of action in a specific situation.

The contrast of withMUSIC minus withoutMUSIC conditions revealed subcortical activation in (1) the amygdala/hippocampus implicated in emotion and learning, (2) the hypothalamus and brainstem implicated in autonomic and endocrine responses, and (3) the anterior cerebellum implicated in behaviour preparation. These findings lend partial support to our hypothesis that music caused an increase in empathic concern and a shift in perspective taking. When the participants were instructed to imagine an emotional scene in the absence of music, they did not take an affective perspective. This is evidenced by the finding that the contrasts of positive and negative withoutMUSIC conditions minus the baseline showed no significant activation. On the other hand, when these imagined scene were accompanied with congruent music that represented the protagonist's emotions, feelings, and emotion-related movements/vocalizations (Molnar-Szakacs & Overy, 2006; Panksepp, 2009), the participants tended to take an affective perspective. From an evolutionary perspective, a precursor of human music may be an affective signaling system common to many socially living animals (Altenmüller, 2013; Snowdon *et al.*, 2015). In human cultures, the emotional power of music is closely linked to its communicative properties, and music is frequently used to facilitate listener's engagement in social functions that support the survival of the individual and the continuation of the species, including cooperation, attachment, and social cohesion (Huron, 2001; Koelsch, 2012; Schulkin & Raglan, 2014; Tarr *et al.*, 2014; Honing *et al.*, 2015). In the present study, the emotional music associated with an imagined scenario may be a salient social signal with meanings concerning approach or avoidance behaviours. It should be borne in mind that scene imagery differs from scene perception in that scene imagery entails the generation, maintenance, and visualization of an environmental setting in which an event can be mentally experienced (Burgess *et al.*, 2001; Hassabis *et al.*, 2007b). Under the guidance of the musical stimuli, the participants seemed to generate emotional responses as if they are the protagonist in the self-constructed emotional scenes. The combination of imagined scenes and emotional music might be more effective for emotion contagion than the combination of films/pictures and emotional music. While this hypothesis awaits validation in prospective studies, it is noteworthy that the present study demonstrated significant activation in the hypothalamus and LC for the contrast of withMUSIC > withoutMUSIC, whereas previous studies mainly reported altered functional connectivity of the hypothalamus and LC in response of music (Lerner *et al.*, 2009; Koelsch & Skouras, 2014).

Another finding in the present study was that a number of cortical regions exhibited significantly greater activity for the positive-emotion than negative-emotion conditions, including the precuneus, supramarginal gyrus, insula, fusiform gyrus, inferior temporal gyrus, and medial orbitofrontal cortex. This activation pattern is strikingly similar to the results of Wassiliwizky et al. (2017), which showed that peak aesthetic pleasure induced by poetry was associated with increased activity in the precuneus, supramarginal gyrus, insula, and fusiform gyrus. Given that readers' mental imagery plays a prominent role in poetry appreciation, it seems plausible that imagery of pleasurable scenes activated brain areas similar to those associated with pleasurable sentences in poetry. The finding of greater activity in the medial orbitofrontal cortex for the positive-emotion than negative-emotion conditions accords well with previous results showing its key role in the evaluation of pleasurable music (Menon & Levitin, 2005; Trost *et al.*, 2012; Salimpoor *et al.*, 2013; Li *et al.*, 2015) and positive social scenarios (Lin *et al.*, 2012).

With regard to future studies on the relationships between music, scene, and emotion, some aspects have to be taken into account. First, the major screening criterion to be able to participate in the fMRI experiment was the score of an online test, through which we enrolled a group having relatively consistent recognized affect contents conveyed by the musical stimuli. It would be of interest in future studies to explore the relationship between visual imagery via music and the ability to recognize emotions conveyed by music, as well as the underlying neural correlates. Second, this study did not record any behavioral data relevant to cognitive functions during the fMRI scan. Future studies could ask participants to rate the vividness of the visual imagery and examine how this rate score correlates to the emotional responses. Third, the final sample of fMRI data consisted of 11 females and 5 males. Future studies with a larger sample size and a good gender balance are needed to verify the current results.

To sum up, the present study used combinations of imagined scenes and affectively congruent music to explore the neural substrates of socioemotional processing. Compared to scene imagery without music, imaging emotional scenes with congruent music evoked greater activity in the bilateral TP and several subcortical areas. The multimodal socioemotional processing in the TP may have modulatory effects on the amygdala and hippocampus, which in turn interact with the hypothalamus, brainstem, and anterior cerebellum. These subcortical responses provide evidence that emotional music congruent with an imagined emotional scene may be a salient social signal, which tends to cause preparation of approach/avoidance behaviors and emotional responses in listeners. Future studies should investigate the factors impacting the therapeutic efficacy of combinations of visual imagery and music listening.

## Acknowledgements

The authors would like to express our gratitude to Prof. Tai-Li Chou and Dr. Andrew Chang for helpful discussion. The authors also thank all the subjects who participated in the study, and to Wei-Jin Lin and Chao-Ju Chen for stimuli preparation and data collection. This study was financially supported by grant projects (MOST 104-2410-H-002-114 and MOST 106-2420-H-002-009) from Ministry of Science and Technology, Taiwan.

## Author Contributions

Chen-Gia Tsai designed the study and oversaw the data collection. Chia-Wei Li and Tzu-Han Cheng analyzed the data. Chen-Gia Tsai, Tzu-Han Cheng, and Chia-Wei Li wrote the paper.

## Conflict of interest

The authors have no conflicts of interest to report.

## References

Altenmüller, E., Kopiez, R., Grewe, O. (2013) A contribution to the evolutionary basis of music: lessons from the chill response. In: Altenmüller, E., Schmidt, S., Zimmermann, E. (Eds.), Evolution of Emotional Communication. *Oxford University Press, Oxford*, 313-335.

Backus, A.R., Bosch, S.E., Ekman, M., Grabovetsky, A.V. & Doeller, C.F. (2016) Mnemonic convergence in the human hippocampus. *Nature communications*, **7**, 11991.

Baumgartner, T., Lutz, K., Schmidt, C.F. & Jancke, L. (2006) The emotional power of music: how music enhances the feeling of affective pictures. *Brain research*, **1075**, 151-164.

Bickart, K.C., Dickerson, B.C. & Barrett, L.F. (2014) The amygdala as a hub in brain networks that support social life. *Neuropsychologia*, **63**, 235-248.

Burgess, N., Becker, S., King, J.A. & O'Keefe, J. (2001) Memory for events and their spatial context: models and experiments. *Philos T R Soc B*, **356**, 1493-1503.

Chen, I.P., Lin, Z.X., Tsai, C.G. (2013) A felt-emotion-based corpora of music emotions (in Chinese). *Chinese Journal of Psychology*, **55(4)**, 571-599.

Damasio, A.R., Grabowski, T.J., Bechara, A., Damasio, H., Ponto, L.L., Parvizi, J. & Hichwa, R.D. (2000) Subcortical and cortical brain activity during the feeling of self-generated emotions. *Nature neuroscience*, **3**, 1049-1056.

Downey, L.E., Blezat, A., Nicholas, J., Omar, R., Golden, H.L., Mahoney, C.J., Crutch, S.J. & Warren, J.D. (2013) Mentalising music in frontotemporal dementia. *Cortex*, **49**, 1844-1855.

Egermann, H. & McAdams, S. (2013) Empathy and Emotional Contagion as a Link between Recognized and Felt Emotions in Music Listening. *Music Percept*, **31**, 139-156.

Eldar, E., Ganor, O., Admon, R., Bleich, A. & Hendler, T. (2007) Feeling the real world: limbic response to music depends on related content. *Cerebral cortex*, **17**, 2828-2840.

Engen, H.G., Kanske, P., Singer, T., 2017. The neural component-process architecture of endogenously generated emotion. Soc Cogn Affect Neurosci 12, 197–211. doi:10.1093/scan/nsw108

Fan, L., Wang, J., Zhang, Y., Han, W., Yu, C. & Jiang, T. (2014) Connectivity-based parcellation of the human temporal pole using diffusion tensor imaging. *Cereb Cortex*, **24**, 3365-3378.

Hassabis, D., Kumaran, D. & Maguire, E.A. (2007a) Using imagination to understand the neural basis of episodic memory. *Journal of Neuroscience*, **27**, 14365-14374.

Hassabis, D., Kumaran, D., Vann, S.D. & Maguire, E.A. (2007b) Patients with hippocampal amnesia cannot imagine new experiences. *Proceedings of the National Academy of Sciences of the United States of America*, **104**, 1726-1731.

Hassabis, D. & Maguire, E.A. (2007) Deconstructing episodic memory with construction. *Trends Cogn Sci*, **11**, 299-306.

Holly, E.N. & Miczek, K.A. (2016) Ventral tegmental area dopamine revisited: effects of acute and repeated stress. *Psychopharmacology*, **233**, 163-186.

Honing, H., ten Cate, C., Peretz, I. & Trehub, S.E. (2015) Without it no music: cognition, biology and evolution of musicality Introduction. *Philos T R Soc B*, **370**, 5-12.

Hsieh, S., Hornberger, M., Piguet, O. & Hodges, J.R. (2012) Brain correlates of musical and facial emotion recognition: evidence from the dementias. *Neuropsychologia*, **50**, 1814-1822.

Hunter, P.G., Schellenberg, E.G. & Schimmack, U. (2010) Feelings and Perceptions of Happiness and Sadness Induced by Music: Similarities, Differences, and Mixed Emotions. Psychol Aesthet Crea, 4, 47-56.

Huron, D. (2001) Is music an evolutionary adaptation? *Ann Ny Acad Sci*, **930**, 43-61.

Josephs, O. & Henson, R.N. (1999) Event-related functional magnetic resonance imaging: modelling, inference and optimization. *Philosophical transactions of the Royal Society of London. Series B, Biological sciences*, **354**, 1215-1228.

Juslin, P.N., Vastfjall, D., 2008. Emotional responses to music: the need to consider underlying mechanisms. Behav Brain Sci 31, 559–75– discussion 575–621.

Karagozoglu, S., Tekyasar, F. & Yilmaz, F.A. (2013) Effects of music therapy and guided visual imagery on chemotherapy-induced anxiety and nausea-vomiting. *Journal of clinical nursing*, **22**, 39-50.

Koelsch, S. (2012) *Brain and music*. Wiley-Blackwell, Chichester, West Sussex ; Hoboken, NJ.

Koelsch, S. & Skouras, S. (2014) Functional centrality of amygdala, striatum and hypothalamus in a "small-world" network underlying joy: an fMRI study with music. *Human brain mapping*, **35**, 3485-3498.

Leigh, R., Oishi, K., Hsu, J., Lindquist, M., Gottesman, R.F., Jarso, S., Crainiceanu, C., Mori, S. & Hillis, A.E. (2013) Acute lesions that impair affective empathy. *Brain : a journal of neurology*, **136**, 2539-2549.

Lerner, Y., Papo, D., Zhdanov, A., Belozersky, L. & Hendler, T. (2009) Eyes wide shut: amygdala mediates eyes-closed effect on emotional experience with music. *PloS one*, **4**, e6230.

Li, C.W., Chen, J.H. & Tsai, C.G. (2015) Listening to music in a risk-reward context: The roles of the temporoparietal junction and the orbitofrontal/insular cortices in reward-anticipation, reward-gain, and reward-loss. *Brain research*, **1629**, 160-170.

Li, S.B., Jones, J.R. & de Lecea, L. (2016) Hypocretins, Neural Systems, Physiology, and Psychiatric Disorders. *Current psychiatry reports*, **18**, 7.

Lin, A., Adolphs, R. & Rangel, A. (2012) Social and monetary reward learning engage overlapping neural substrates. *Soc Cogn Affect Neurosci*, **7**, 274-281.

Lundqvist, L.O., Carlsson, F., Hilmersson, P. & Juslin, P.N. (2009) Emotional responses to music: experience, expression, and physiology. Psychol Music, 37, 61-90.

McKinney, C.H. & Honig, T.J. (2017) Health Outcomes of a Series of Bonny Method of Guided Imagery and Music Sessions: A Systematic Review. *Journal of music therapy*, **54**, 1-34.

Mears, D. & Pollard, H.B. (2016) Network science and the human brain: Using graph theory to understand the brain and one of its hubs, the amygdala, in health and disease. *Journal of neuroscience research*, **94**, 590-605.

Mehrabian, A. & Epstein, N., 1972. A measure of emotional empathy. J Pers 40, 525–543.

Menon, V. & Levitin, D.J. (2005) The rewards of music listening: response and physiological connectivity of the mesolimbic system. *NeuroImage*, **28**, 175-184.

Miu, A.C. & Baltes, F.R. (2012) Empathy Manipulation Impacts Music-Induced Emotions: A Psychophysiological Study on Opera. PloS one, 7.

Molnar-Szakacs, I. & Overy, K. (2006) Music and mirror neurons: from motion to 'e'motion. *Soc Cogn Affect Neurosci*, **1**, 235-241.

Murphy, C., Rueschemeyer, S.A., Watson, D., Karapanagiotidis, T., Smallwood, J. & Jefferies, E. (2017) Fractionating the anterior temporal lobe: MVPA reveals differential responses to input and conceptual modality. *Neuroimage*, **147**, 19-31.

Olson, I.R., McCoy, D., Klobusicky, E. & Ross, L.A. (2013) Social cognition and the anterior temporal lobes: a review and theoretical framework. *Soc Cogn Affect Neurosci*, **8**, 123-133.

Omar, R., Henley, S.M., Bartlett, J.W., Hailstone, J.C., Gordon, E., Sauter, D.A., Frost, C., Scott, S.K. & Warren, J.D. (2011) The structural neuroanatomy of music emotion recognition: evidence from frontotemporal lobar degeneration. *NeuroImage*, **56**, 1814-1821.

Panksepp, J. (2009) The emotional antecedents to the evolution of music and language. *Music Sci*, 229-259.

Pehrs, C., Deserno, L., Bakels, J.H., Schlochtermeier, L.H., Kappelhoff, H., Jacobs, A.M., Fritz, T.H., Koelsch, S. & Kuchinke, L. (2014) How music alters a kiss: superior temporal gyrus controls fusiform-amygdalar effective connectivity. *Social cognitive and affective neuroscience*, **9**, 1770-1778.

Pehrs, C., Zaki, J., Schlochtermeier, L.H., Jacobs, A.M., Kuchinke, L. & Koelsch, S. (2017) The Temporal Pole Top-Down Modulates the Ventral Visual Stream During Social Cognition. *Cerebral cortex*, **27**, 777-792.

Robinson, J.L., Laird, A.R., Glahn, D.C., Blangero, J., Sanghera, M.K., Pessoa, L., Fox, P.M., Uecker, A., Friehs, G., Young, K.A., Griffin, J.L., Lovallo, W.R. & Fox, P.T. (2012) The functional connectivity of the human caudate: an application of meta-analytic connectivity modeling with behavioral filtering. *NeuroImage*, **60**, 117-129.

Royet, J.P., Zald, D., Versace, R., Costes, N., Lavenne, F., Koenig, O. & Gervais, R. (2000) Emotional responses to pleasant and unpleasant olfactory, visual, and auditory stimuli: a positron emission tomography study. *The Journal of neuroscience : the official journal of the Society for Neuroscience*, **20**, 7752-7759.

Salimpoor, V.N., van den Bosch, I., Kovacevic, N., McIntosh, A.R., Dagher, A. & Zatorre, R.J. (2013) Interactions between the nucleus accumbens and auditory cortices predict music reward value. *Science*, **340**, 216-219.

Schulkin, J. & Raglan, G.B. (2014) The evolution of music and human social capability. *Frontiers in neuroscience*, **8**.

Snowdon, C.T., Zimmermann, E. & Altenmuller, E. (2015) Music evolution and neuroscience. *Progress in brain research*, **217**, 17-34.

Tarr, B., Launay, J. & Dunbar, R.I.M. (2014) Music and social bonding: "self-other" merging and neurohorrnonal mechanisms. *Frontiers in psychology*, **5**.

Trost, W., Ethofer, T., Zentner, M. & Vuilleumier, P. (2012) Mapping aesthetic musical emotions in the brain. *Cereb Cortex*, **22**, 2769-2783.

Ulrich-Lai, Y.M. & Herman, J.P. (2009) Neural regulation of endocrine and autonomic stress responses. *Nature reviews. Neuroscience*, **10**, 397-409.

Vuoskoski, J.K. & Eerola, T. (2015) Extramusical information contributes to emotions induced by music. *Psychol Music*, **43**, 262-274.

Wassiliwizky, E., Koelsch, S., Wagner, V., Jacobsen, T. & Menninghaus, W. (2017) The emotional power of poetry: neural circuitry, psychophysiology and compositional principles. *Soc Cogn Affect Neurosci*, **12**, 1229-1240.

Waterhouse, B.D. & Navarra, R.L., 2019. The locus coeruleus-norepinephrine system and sensory signal processing: A historical review and current perspectives. Brain Res. 1709, 1–15.

Table 1. Linguistic stimuli used for guiding scene imagery.

| | |
|---|---|
| **Positive Emotion** | A hippo and a puppy are enjoying a stroll together. They walk up a hill, slide down the slope, and continue on their way. |
| | A piglet and a duckling are playing around and poking each other. After letting out a few farts, they set out on a journey. |
| | A blonde has finished her shopping, and is humming a song as she walks down the street. |
| | A tipsy old fool can't keep straight after having too much wine, and bumbles around the stage with his comedic antics. |
| | A clown peeps out from the side of the stage, then swiftly dances onto the stage with a smile on his face. |
| **Negative Emotion** | A young revolutionary is sentenced to death, and cries out in anguish in protest of the injustice. |
| | A woman is sobbing in despair, her voice trembling, yet is resolute in accusing the offender. |
| | A person walks through a misty ghost town (hell), as evil spirits suddenly appear on all sides. |
| | A girl has realized that her romantic relationship has ended after many years, and tears trickle down her face. |
| | A hero returns to his hometown, only to see total destruction. As he holds the body of a dear relative, he turns his agonized face toward the heavens in an empty glaze. |
| **Neutral Emotion** | Several sparrows are hopping along the balcony, chattering among themselves. |
| | A few frogs are swimming slowly at the bottom of the pond, and eventually stop at its edge. |
| | In an elementary school classroom, a person is wiping the windows, while another is hosing and mopping the floor. |
| | As a metro train pulls into a station, some rush off onto the platform, while others hurry inside. |
| | On a school's racing track, some are jogging, others are strolling along. |

Table 2. Montreal Neurological Institute (MNI) coordinates and *t*-values for significantly activated brain regions in the contrasts of four experimental conditions minus the baseline (FDR-corrected *p* < 0.05, at least 10 contiguous voxels).

| Volume Information | X | Y | Z | *t*-value | Cluster size (voxel) |
|---|---|---|---|---|---|
| **Positive withMUSIC > Baseline** | | | | | |
| Transverse temporal gyrus | 44 | -20 | 6 | 17.88 | 2035 |
| Superior temporal gyrus | 50 | -20 | 4 | 13.29 | |
| Supramarginal gyrus, insula | 48 | -32 | 16 | 7.70 | |
| Superior temporal pole | 46 | 4 | -14 | 6.34 | |
| Insula | 28 | -30 | 18 | 4.27 | |
| Superior temporal gyrus | -46 | -8 | -6 | 10.11 | 2289 |
| Transverse temporal gyrus, insula | -42 | -22 | 6 | 9.43 | |
| Superior temporal pole | -38 | -2 | -18 | 4.54 | |
| Cerebellar lobule VI | 26 | -58 | -24 | 7.20 | 85 |
| Cerebellar lobule VI | -24 | -60 | -24 | 4.94 | 47 |
| Thalamus | 2 | -20 | 8 | 6.34 | 29 |
| Cerebellar lobule III-V, locus coeruleus | -10 | -36 | -20 | 6.27 | 202 |
| Parahippocampal gyrus, hippocampus, amygdala, hypothalamus | -6 | -6 | -12 | 5.52 | 108 |
| Parahippocampal gyrus | 12 | -12 | -18 | 5.13 | 26 |
| Putamen | 24 | 2 | 4 | 5.06 | 22 |
| Putamen | -24 | -8 | 10 | 4.33 | 23 |
| Temporal pole | -34 | 10 | -24 | 4.78 | 35 |
| Premotor cortex | 48 | -4 | 42 | 4.76 | 15 |
| Premotor cortex | -48 | -2 | 42 | 4.01 | 10 |
| **Positive withoutMUSIC > Baseline** | | | | | |
| No differential activations | | | | | |
| **Negative withMUSIC > Baseline** | | | | | |
| Superior temporal gyrus | 46 | -18 | 6 | 11.76 | 1301 |
| Transverse temporal gyrus | 40 | -26 | 10 | 8.24 | |
| Insula | 54 | -34 | 18 | 6.67 | |
| Superior temporal pole | 46 | 4 | -12 | 5.71 | |
| Superior temporal gyrus | -48 | -18 | 0 | 10.87 | 1909 |
| Superior temporal pole | -52 | 8 | -6 | 9.64 | |
| Transverse temporal gyrus | -42 | -22 | 8 | 9.59 | |
| Middle temporal gyrus | -42 | -26 | 0 | 7.97 | |
| Supramarginal gyrus | -64 | -26 | 14 | 7.29 | |
| Caudate | -8 | 10 | 18 | 7.09 | 66 |
| | -8 | 18 | 8 | 4.53 | 14 |
| | -14 | -8 | 24 | 4.44 | 10 |
| Temporal pole | 42 | 8 | -24 | 6.97 | 23 |
| Hippocampus, parahippocampal gyrus, amygdala, midbrain | -20 | -14 | -20 | 6.16 | 110 |
| | 16 | -6 | -10 | 4.88 | 10 |
| Cerebellar lobule III-V, locus coeruleus | -4 | -36 | 0 | 5.98 | 146 |
| | 12 | -32 | -20 | 5.37 | 51 |
| Ventral tegmental area | -2 | -20 | -16 | 5.48 | 34 |
| Cerebellar lobule VI | 32 | -66 | -24 | 5.42 | 51 |
| Hypothalamus | -4 | -6 | -12 | 5.38 | 34 |
| Premotor cortex | -50 | 4 | 50 | 5.34 | 41 |
| Thalamus | 4 | -6 | -4 | 5.17 | 55 |
| Cingulate gyrus | 10 | -8 | 28 | 4.63 | 45 |
| Putamen | -26 | 6 | 14 | 4.26 | 11 |
| | -22 | -4 | 14 | 4.23 | 14 |
| **Negative withoutMUSIC > Baseline** | | | | | |
| No differential activations | | | | | |

Table 3. ANOVA results (FDR-corrected *p* < 0.05, at least 10 contiguous voxels). There was no significant difference for withoutMUSIC > withMUSIC and for Negative Emotion > Positive Emotion.

| Volume Information | X | Y | Z | t-value | Cluster size (voxel) |
|---|---|---|---|---|---|
| **Main effect of music:   withMUSIC > withoutMUSIC** | | | | | |
| Transverse temporal gyrus (L) | 40 | -28 | 8 | 14.77 | 2303 |
| Middle temporal gyrus (L) | 54 | -8 | 2 | 14.42 | |
| Superior temporal gyrus, insula (L) | 48 | -10 | 2 | 13.54 | |
| Superior temporal pole (L) | 46 | 4 | -14 | 7.43 | |
| Superior temporal gyrus, supramarginal gyrus (L) | 62 | -36 | 12 | 6.04 | |
| Superior temporal gyrus (R) | -50 | -14 | 0 | 14.66 | 2616 |
| Transverse temporal gyrus (R) | -52 | -24 | 8 | 12.82 | |
| Rolandic operculum (R) | -56 | -18 | 4 | 12.77 | |
| Supramarginal gyrus (R) | -60 | -24 | 12 | 12.51 | |
| Insula (R) | -46 | -2 | -12 | 11.58 | |
| Middle temporal gyrus (R) | -60 | -14 | -2 | 10.28 | |
| Superior temporal pole (R) | -58 | 2 | -12 | 7.20 | |
| Cerebellar lobule III-V | 0 | -40 | -6 | 5.94 | 308 |
| Premotor cortex | -48 | 4 | 52 | 5.39 | 53 |
| Cerebellar lobule VI | 32 | -66 | -24 | 4.76 | 105 |
| Cerebellar lobule VI | -26 | -60 | -26 | 3.69 | 54 |
| Amygdala, hippocampus, parahippocampal gyrus, hypothalamus | -16 | -6 | -18 | 3.46 | 147 |
| Ventral tegmental area | -2 | -16 | -14 | 2.70 | 12 |
| Thalamus | 0 | -12 | 8 | 2.36 | 56 |
| **Main effect of emotion valence: Positive emotion > Negative emotion** | | | | | |
| Superior temporal gyrus (R) | 50 | -20 | 4 | 5.92 | 1465 |
| Transverse temporal gyrus (R) | 46 | -18 | 6 | 5.49 | |
| Insula (R) | 42 | -18 | 0 | 5.14 | |
| Postcentral gyrus (R) | 66 | -24 | 14 | 3.50 | |
| Rolandic operculum (R) | 58 | -6 | 10 | 2.78 | |
| Superior temporal gyrus (L) | -48 | -14 | 0 | 4.06 | 1095 |
| Transverse temporal gyrus (L) | -40 | -22 | 6 | 3.81 | |
| Supramarginal gyrus (L) | -66 | -32 | 14 | 3.47 | |
| Middle temporal gyrus (L) | -64 | -12 | -6 | 3.26 | |
| Superior occipital gyrus | -20 | -58 | 18 | 4.92 | 355 |
| Cuneus | -14 | -64 | 24 | 5.15 | |
| Precuneus/posterior cingulate gyrus | -24 | -64 | 22 | 3.91 | |
| Precuneus/posterior cingulate gyrus | 16 | -68 | 30 | 3.53 | 31 |
| | 16 | -50 | 14 | 3.07 | 72 |
| Middle/inferior temporal gyrus, middle occipital gyrus | 50 | -72 | -6 | 4.59 | 240 |
| | -54 | -62 | 0 | 3.95 | 70 |
| Superior/middle occipital gyrus, precuneus | 30 | -70 | 36 | 4.56 | 89 |
| Superior/inferior parietal gyrus | -28 | -62 | 52 | 4.56 | 30 |
| | 14 | -70 | 56 | 3.28 | 12 |
| Inferior parietal gyrus, supramarginal gyrus, postcentral gyrus | -44 | -52 | 44 | 3.14 | 57 |
| | -46 | -42 | 50 | 3.48 | 40 |
| | -54 | -26 | 46 | 3.45 | 86 |
| Supramarginal gyrus, inferior parietal gyrus, postcentral gyrus, precentral gyrus | 64 | -24 | 32 | 4.48 | 307 |
| Fusiform gyrus, parahippocampal gyrus, lingual gyrus | -24 | -46 | -14 | 4.45 | 75 |
| | 38 | -42 | -22 | 3.38 | 18 |
| Precentral gyrus, postcentral gyrus | -34 | -24 | 58 | 4.08 | 56 |
| | 44 | -18 | 62 | 3.06 | 19 |
| Medial orbitofrontal gyrus, cuneus | -10 | 38 | -14 | 4.02 | 31 |
| Precuneus, precentral gyrus, superior parietal gyrus | -10 | -74 | 46 | 3.92 | 105 |
| | -10 | -44 | 56 | 3.03 | 55 |
| Paracentral gyrus, middle cingulate gyrus, supplementary motor area | 10 | -34 | 58 | 3.71 | 58 |
| | -2 | -20 | 54 | 3.08 | 50 |
| Temporal pole | -30 | 6 | -24 | 3.69 | 20 |
| Postcentral gyrus, precentral gyrus, supramarginal gyrus | -48 | -16 | 48 | 3.69 | 119 |
| Superior/middle frontal gyrus, medial orbitofrontal gyrus | -28 | 6 | 58 | 3.21 | 31 |
| | 28 | 2 | 58 | 2.98 | 19 |
| Middle cingulate gyrus | -6 | -28 | 44 | 2.81 | 10 |
| **Interaction effect:** Positive-emotion withMUSIC - Negative-emotion withMUSIC ≠ Positive-emotion withoutMUSIC - Negative-emotion withoutMUSIC | | | | | |
| Superior temporal gyrus, insula | 52 | -18 | 4 | 30.82 | 43 |
| | -40 | -26 | 2 | 24.18 | 21 |
| | 40 | -28 | 8 | 19.14 | 45 |
| | 62 | -22 | 6 | 15.11 | 15 |
| Ventral tegmental area | 2 | -22 | -18 | 29.77 | 12 |

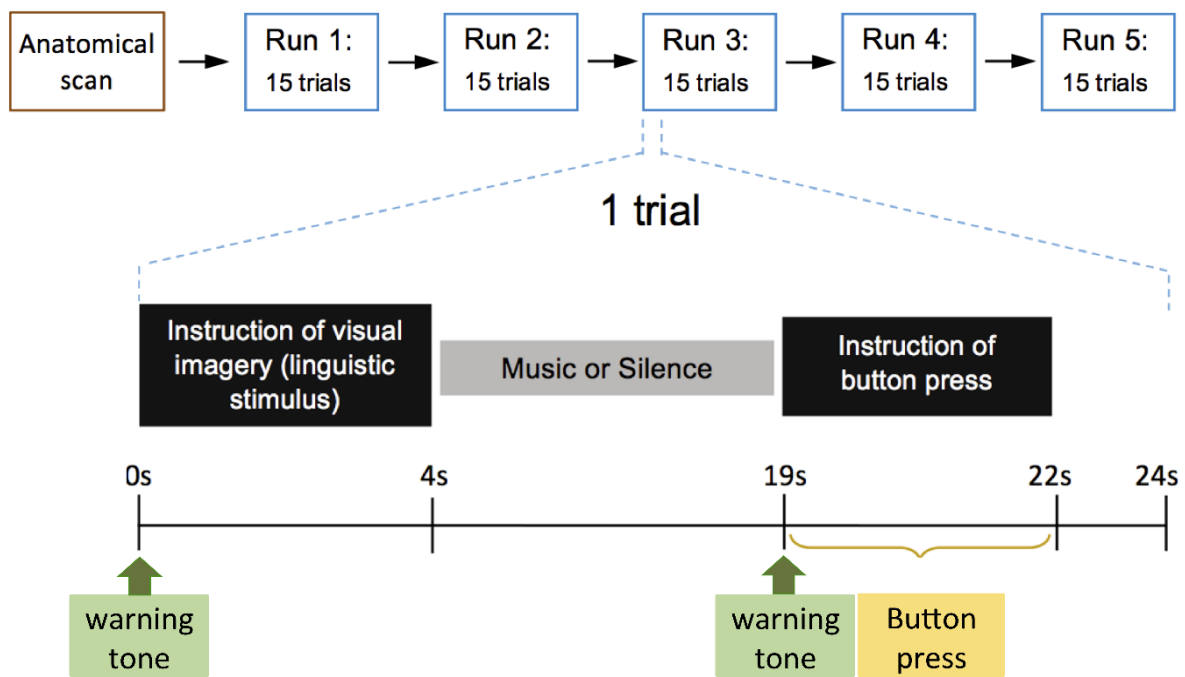

Figure 1. Schematic representation of the fMRI procedure. There were five runs in fMRI scanning session. Each run contained 15 trials with a duration of 24 seconds. In each trial, the participants were instructed to visually imagine the scene described by the linguistic stimulus shown on the screen. Meanwhile, a musical stimulus or silence was presented. At the end of each trial, an instruction was presented on the screen, which asked the participant to report their present subjective emotion valence.

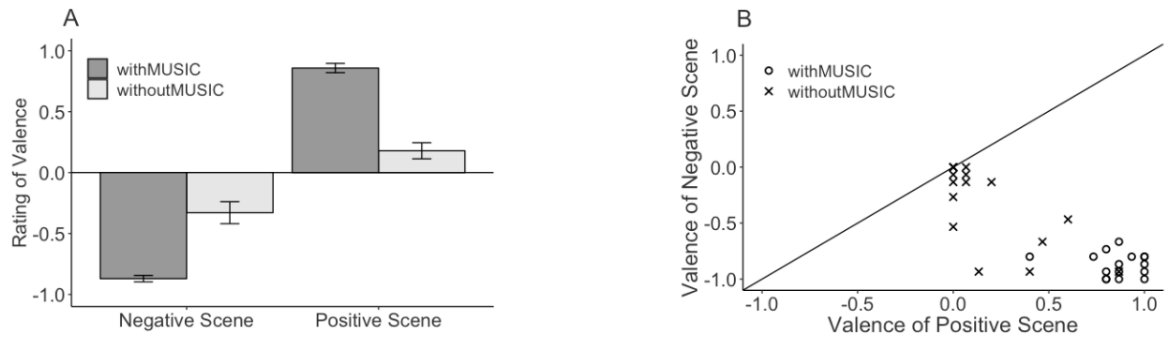

Figure 2. Behavioural results of the self-rating scores of subjective emotion valence. (A) Bar graph with mean and standard error of the mean (SEM). (B) Scatterplot revealing the relationship between the self-rating scores of subjective emotion valence of positive scenes (*x* axis) and those of negative scenes (*y* axis). The diagonal line has slope 1 and intercept 0. The data points are below the diagonal line and gathering on the bottom right. This suggests the major trends of (1) rating higher subjective emotion valence for the positive scenes, and (2) rating lower subjective emotion valence for the negative scenes. A majority of the data points for the withoutMUSIC condition are close to the origin (0,0), while a majority of the data points for the withMUSIC condition are close to (1,-1). This suggests that music considerably intensified the subjective emotions.

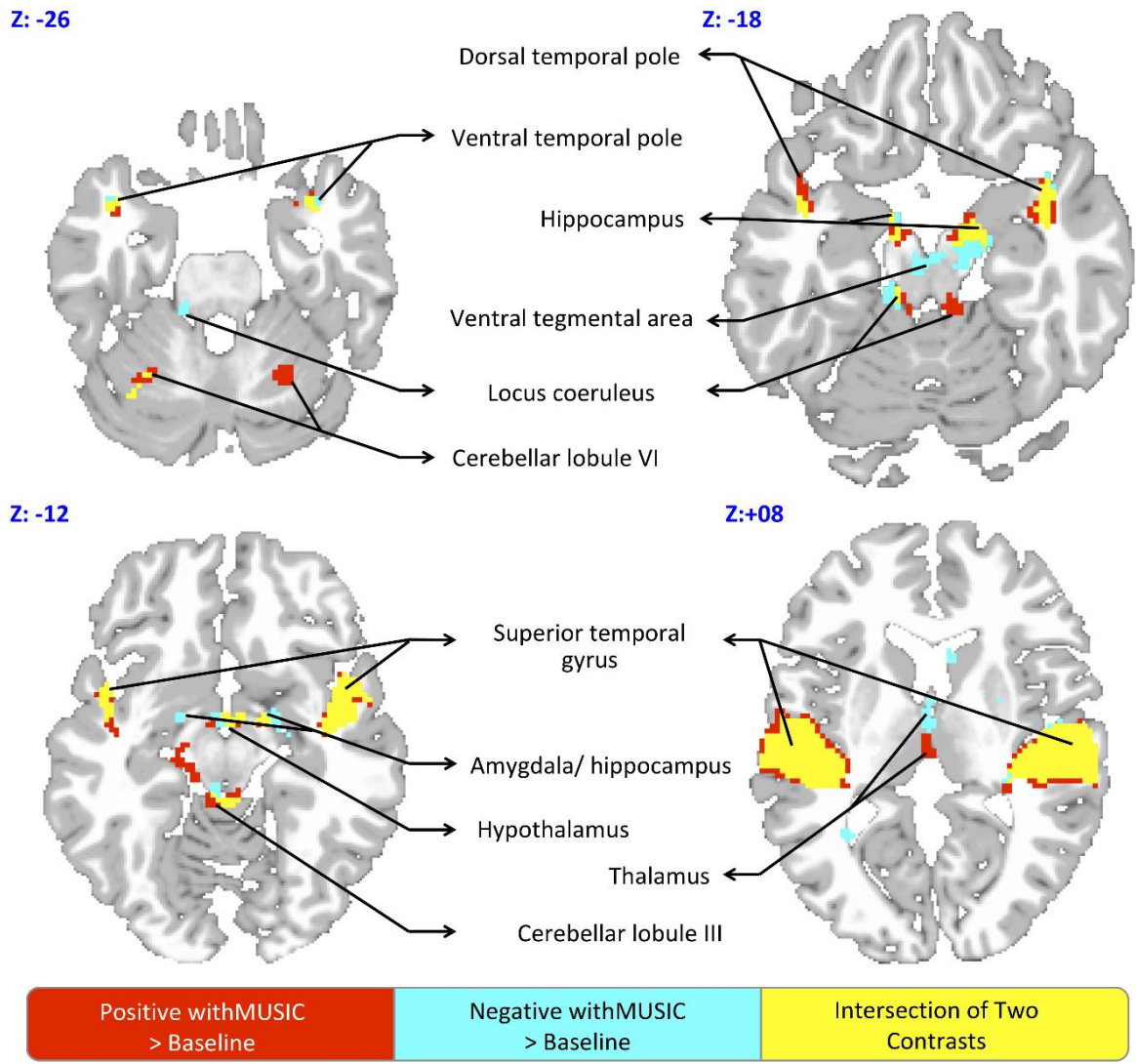

Figure 3. Contrasts of the positive emotion withMUSIC condition minus baseline and the negative emotion withMUSIC condition minus baseline (FDR-corrected $p < 0.05$, at least 10 contiguous voxels).

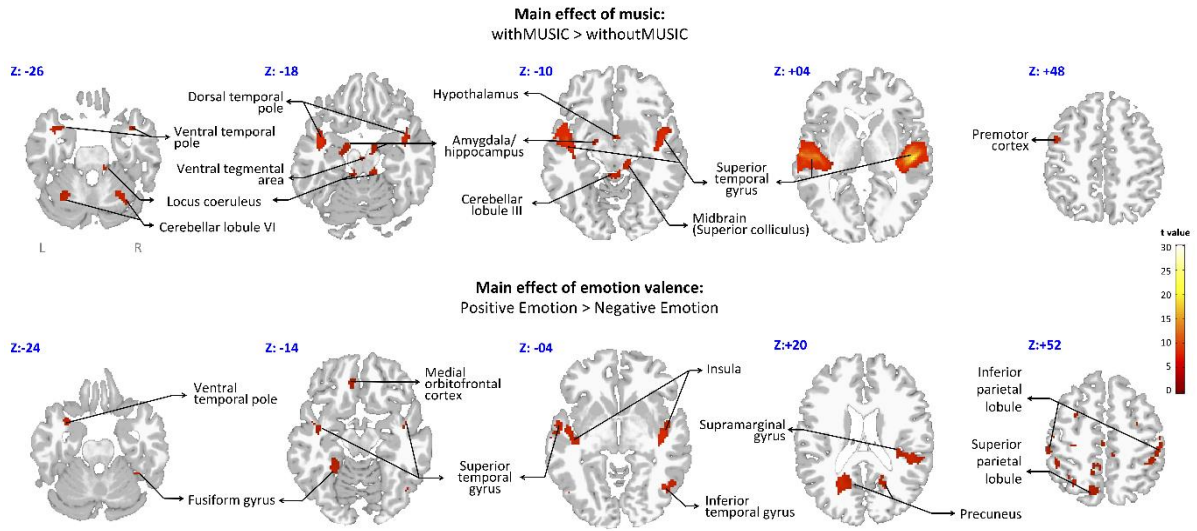

Figure 4. Main effects of music and emotion valence (FDR-corrected *p* < 0.05, at least 10 contiguous voxels).

## Interaction Effect of ANOVA

Positive emotion valence with music - Positive emotion valence without music
≠
Negative emotion valence with music - Negative emotion valence without music

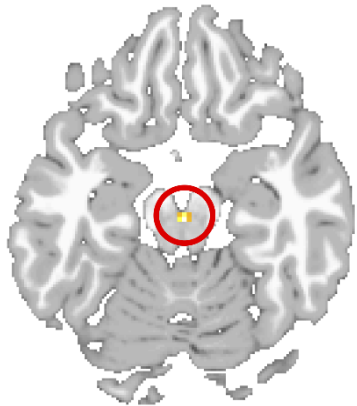
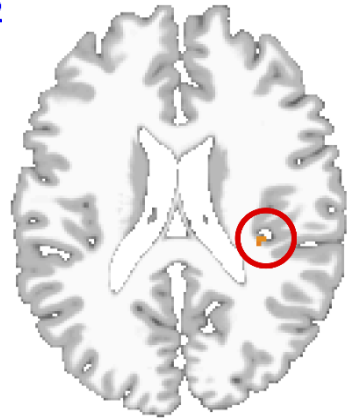
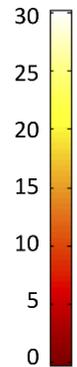
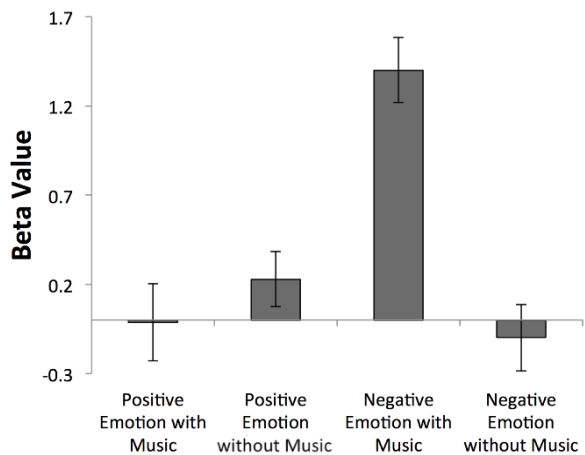
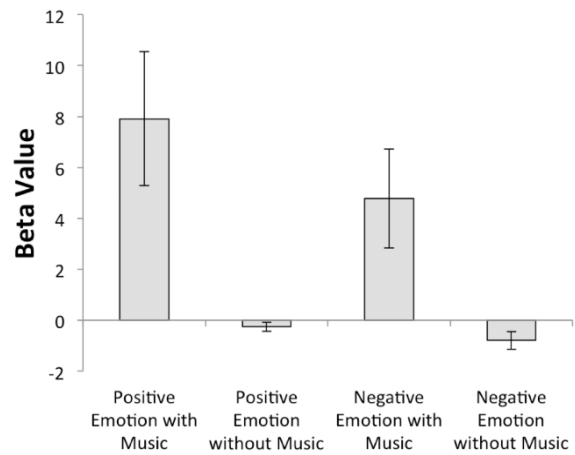

Figure 5. Interaction effect between emotion valence and music. The beta-values of VTA and posterior insula for the four experimental conditions were demonstrated. For the withMUSIC conditions, positive emotions were associated with increased activation in the posterior insula and decreased activation in the posterior VTA compared to negative emotions.